\documentclass[a4paper,floatfix,amsmath,amssymb,prb,twocolumn,subeqn]{revtex4}
\usepackage{amsmath}
\usepackage{graphicx}
\usepackage{epsf}
\usepackage{dcolumn}
\usepackage{bm} 
\begin{document}
\title {Monte Carlo simulation of joint density of states of two 
continuous spin models using Wang-Landau-Transition-Matrix Algorithm\\
}
\author{Shyamal Bhar}
\email{sbhar@research.jdvu.ac.in }
\affiliation{Department of Physics, Jadavpur University, Kolkata-700032, India.}
\author{ Soumen Kumar Roy}
\email{skroy@phys.jdvu.ac.in }
\affiliation{Department of Physics, Jadavpur University, Kolkata-700032, India.}
\begin {abstract}
\noindent Monte Carlo simulation has been performed in one-dimensional
Lebwohl-Lasher model and two dimensional XY-model using the Wang-Landau
and the Wang-Landau-Transition-Matrix Monte Carlo methods. Random walk
has been performed in the two-dimensional space comprising of energy-order
parameter and energy-correlation function and the joint density of states
(JDOS) were obtained. From the JDOS the order parameter, susceptibility
and correlation function are calculated. Agreement between the results
obtained from the two algorithms is very good.\\

\noindent PACS: 64.60.De; 61.30.-v; 05.10.Ln\\
\noindent Keywords: Wang Landau, Transition Matrix. Joint Density of states
\end{abstract}
\maketitle
\textheight=8.9in
\textheight=8.9in
\section {\bf Introduction}
\noindent During the last couple of years or so a number of Monte-Carlo
(MC) algorithms have been proposed which directly determine the density of
states (DOS) of a system. One of these is the Transition Matrix Monte Carlo
(TMMC) algorithm developed by Oliveria {\it et al} \cite{oliveria1} and 
subsequently
generalized by Wang and co-workers \cite{swendsen}. In this algorithm, during a
random walk in the energy space, one keeps a record of the transitions
between the microstates of the system. The entire history of the transitions
is then used to obtain the density of states of the system. More recently
Wang and Landau \cite{wlprl} proposed another algorithm which goes by 
their name (WL)
and has drawn wide attention of investigators. This algorithm employs a method
of flat histograms, while a random walk is performed in the energy space and
estimates the DOS of the system by using an iterative scheme. Both methods,
the TMMC and WL, depend on broad sampling of the phase space and are easy to
implement. A knowledge of the DOS of a system as a function of energy,
$\Omega(E)$ enables one to calculate the partition function $Z$ by a simple
Boltzmann reweighting: $Z(\beta)={\displaystyle\sum_E{\Omega(E)e^{-\beta E}}}$,
where $\beta$ is the inverse temperature $1/T$. With a knowledge of the 
partition function one can calculate the averages of thermodynamic quantities
which are directly related to energy. It has been established that while the TMMC method
gives more accurate estimation of the DOS, the WL algorithm is more efficient
in sampling the phase space.\\
\noindent Shell { \it et al} proposed \cite{shell} an algorithm which is an amalgamation
of the two algorithms and utilizes the benefits of each. This algorithm, now
known as the Wang Landau Transition Matrix (WLTM) Monte Carlo algorithm, 
is at the same
time efficient and accurate. Shell {\it et al} applied the algorithm to a
two-dimensional Ising model and a Lennard-Jones fluid. More recently
Ghulghazarya {\it et al} \cite{ghulgha} have applied this method to simulate protein and 
peptide.\\
\noindent The sampling of the phase space which is done in each of the
above methods need not be restricted to the evaluation of the density
of states as a function of energy alone. One can determine the DOS or to be
more specific, the Joint Density of States (JDOS) with a substantially more
book-keeping. The JDOS $\Omega(E, \phi)$ is a function of some variable
$\phi$ (which can be an order parameter, spin-spin correlation or any other
observable) besides the energy $E$. This partition function is determined from:
$Z(\beta)={\displaystyle\sum_E \sum_{\phi}{\Omega(E, \phi)e^{-\beta E}}}$.
The ensemble average of any function of $\phi$ at an inverse temperature 
$\beta$ is then given by
\begin{equation}\label{ave_fun}
<f(\phi, \beta)>=\frac{{\displaystyle\sum_E \sum_{\phi} f(\phi) \Omega(E, \phi)e^{-\beta E}}}{{\displaystyle\sum_E \sum_{\phi} \Omega(E, \phi) e^{-\beta E}}}
\end{equation}
\noindent Most of the investigators have so far worked on the determination
of DOS or JDOS in discrete systems using the three above mention algorithms
\cite{shell2, yan, landau, xu, kisor}. Even in this domain the amount of work 
reported on the determination
of JDOS is relatively small. In an earlier paper \cite{shyamal} we have reported
on the working of WLTM in two continuous lattice-spin models. While the
discrete systems like Ising or Potts model can be handled in a straight
forward manner, the investigation of continuous systems are more tedious.
The range of energy (and another observable in case of two-dimensional random
walk) needs to be discretized and several parameters appear which are to be
chosen properly for the determination of DOS or JDOS. In the present 
communication we report the determination of JDOS using the WLTM algorithm
in two continuous lattice spin models. The quantities evaluated other than
energy and specific heat are the order parameter, susceptibility and 
correlation function. One of the models is the one-dimensional Lebwohl-Lasher
model \cite{ll}, which is exactly solvable \cite{romerio} and therefore 
allows us to check the
accuracy of the results of our simulation. The other system 
is the two-dimensional
XY-model where exact solutions are not available and we have compared the
results with those obtained from the JDOS determined using the WL algorithm.
The aim of the present work is to test the feasibility of the determination
of JDOS in continuous models using the WLTM algorithm. This turned out to be a
some what difficult task as a formidable amount of computer memory is necessary
even for systems of moderate size.
\section {\bf The different algorithms}
\subsection {\bf The Wang Landau algorithm}
\noindent We outline below the method of determination of the JDOS, 
$\Omega(E, \phi)$ using the WL algorithm. In a system where $E$ and $\phi$
are continuous variables, discretization in required to label the macrostates
of the system. The ranges of $E$ and $\phi$ are divided into a large number
of bins of width $d_e$ and $d_{\phi}$ respectively. Let $E_I$ and $\phi_J$
be the mean energy and mean order parameter (or correlation function) 
corresponding to the $I^{th}$ bin of energy and the $J^{th}$ bin of the
order parameter (correlation function) respectively. Then $\Omega(I, J)$
denotes the number of microstates of the system having energy $E_I$ and
order parameter (correlation function) $\phi_J$ or simply the degeneracy
of the macrostate $(I,J)$. Since $\Omega(I, J)$ is a very large number,
it is convenient to work with its natural logarithm and we use 
$g(I, J)=ln \Omega(I,J)$.\\

\noindent We perform two dimensional random walk in the 
$(E-\phi)$ space. Initially we do not
have any knowledge of $g(I,J)$, and set $g(I,J)=0$ for all values of $I$ and
$J$. Also, an histogram count $H(I,J)$ of the states visited during the
random walk is maintained. 
The WL algorithm generates the JDOS profile, which
progressively approaches the actual density of states of the system. The
algorithm starts with some microstate of the system and successive 
microstates are generated
by rotating one spin at a time. Let $(I,J)$ and $(K,L)$ be the macrostates 
before and after
rotating the spin and the corresponding microstates are $(i,j)$ and $(k,l)$,
where $i\in I, j\in J$ and $k\in K, l\in L$. The transition probability
from state $(i,j)$ to $(k,l)$ is given by
\begin{equation}\label{aprob}
P(i,j \rightarrow k,l) = min \left(\frac{\Omega(I,J)}{\Omega(K,L)},1 \right)
\end{equation}
\noindent Thus the acceptance probability of the new state is inversely
proportional to the current density of states. When the new state is
accepted the density of states $g(K,L)$ and the histogram count $H(K,L)$ of
the state $(K,L)$ are modified as
\begin{subequations}
\begin{equation}
g(K,L)=g(K,L)+ln f
\end{equation} 
\begin{equation}
\text{and} \qquad H(K,L)=H(K,L)+1   
\end{equation} 
\end{subequations}
and when the new state is not accepted the old density of state $g(I,J)$
and histogram count $H(I,J)$ are modified as
\begin{subequations}
\begin{equation}
g(I,J)=g(I,J)+ln f
\end{equation}
\begin{equation}
\text{and} \qquad H(I,J)=H(I,J)+1
\end{equation}
\end{subequations}
\noindent Here $f$ is a modification factor whose initial value was chosen
to be equal to $e$. When the histogram is
sufficiently flat (say, 80\%) i.e., histogram count $H(I,J)$ of each
bin $(I,J)$
is at least $80\%$ of the mean histogram $H_M=\left(\sum^{M_{1}}_{I}\sum^{M_2}_{J} H(I,J)\right)/M$
, $M$ being the total number of bins then one iteration is said to be
complete. Then the histogram is reset to zero for all values of $I$ and $J$
and the modification factor $f$ is reduced in some prescribed manner
(we use $lnf \rightarrow lnf/2$). A fresh iteration is started with the modified
value of $lnf$ and the old values of $g(I,J)'s$ which were calculated in the
previous iteration. One continues iterations with the same procedure until the
modification factor becomes sufficiently small (say, $10^{-8}$).
The error introduced in the
joint density of states has been predicted to be proportional to $\sqrt(ln f)$
as is apparent from the theoretical work of Zhou and Bhatt \cite{zhou}. 
This has been
tested for a number of discrete and continuous models and the prediction has
been found to be correct \cite{lee}.\\

\subsection{\bf The Transition Matrix Monte Carlo algorithm}
\noindent  The TMMC algorithm is an efficient algorithm in which one directly
calculates the density of states and was first proposed by Oliveria et al. in
the year 1996. If the transition probability from a microstate $(i,j)$ to
another microstate $(k,l)$ is $t(i,j;k,l)$ and that from a macrostate $(I,J)$
to a macrostate $(K,L)$ is $T(I,J;K,L)$ then
\begin {equation}
T(I,J;K,L)=\frac{1}{\Omega(I,J)}\sum_{i\in I}\sum_{j\in J}
\sum_{k\in K}\sum_{l\in L}{t(i,j;k,l)}
\end {equation}
with the following conditions:
\begin{subequations}
\begin{equation}
\sum_{k,l}{t(i,j;k,l)}=1, \qquad t(i,j;k,l)\ge 0
\end{equation}
\begin{equation}
\text{and}\qquad  \sum_{K,L}{T(I,J;K,L)}=1, \qquad T(I,J;K,L)\ge 0
\end{equation}
\end{subequations}
\noindent If $T(K,L;I,J)$ be the reverse transition then one can write

\begin{equation}\label{tmr1}
\frac{T(I,J;K,L)}{T(K,L;I,J)}=\frac{\Omega(K,L)}{\Omega(I,J)}
\frac{\displaystyle{\sum_{i\in I}\sum_{j\in J}\sum_{k\in K}\sum_{l\in L}
{t(i,j;k,l)}}}{\displaystyle
{\sum_{k\in K}\sum_{l\in L}\sum_{i\in I}\sum_{j\in J}{t(k,l;i,j)}}}
\end{equation}
\noindent The transition probability $t(i,j;k,l)$ actually is a product of two
probabilities,
\begin {equation}\label{tap}
t(i,j;k,l)=a(i,j;k,l)P(i,j;k,l)
\end {equation}
\noindent Here $a(i,j;k,l)$ is the probability of the move $(i,j)\rightarrow 
(k,l)$ being proposed and depends
on the type of the Monte Carlo moves and while $P(i,j;k,l)$ is the probability
of proposed move being accepted and  depends on the configurations
$(i,j)$ and $(k,l)$. One can choose any value of $P$ and an infinite temperature
transition probability $(T_{\infty})$ can be chosen for which $P(i,j;k,l)=1 $ 
for all i,j and k,l. This is particularly easy to understand if one considers
the Metropolis algorithm at infinite temperature.
This process does not affect the JDOS to be determined since it is independent
of temperature.
Using
equations (\ref{tmr1}) and (\ref{tap}) we can write:
\begin{equation}\label{tmr2}
\frac{T_{\infty}(I,J;K,L)}{T_{\infty}(K,L;I,J)}=\frac{\Omega(K,L)}{\Omega(I,J)}
\frac{\displaystyle{\sum_{i\in I}\sum_{j\in J}\sum_{k\in K}\sum_{l\in L}a(i,j;k,l)}}
{\displaystyle{\sum_{k\in K}\sum_{l\in L}\sum_{i\in I}\sum_{j\in J}a(k,l;i,j)}}
\end{equation}
\noindent Again for symmetric moves (single spin flip dynamics)
$a(i,j;k,l)=a(k,l;i,j)$ and the summation terms on right
hand side of equation (\ref{tmr2}) drops out and the equation can be simplified
as

\begin{equation}\label{tw}
\frac{T_{\infty}(I,J;K,L)}{T_{\infty}(K,L;I,J)}=\frac{\Omega(K,L)}{\Omega(I,J)}
\end{equation}
\noindent This equation relates the JDOS with the infinite
temperature transition probabilities. Thus from the knowledge of infinite
temperature transition probabilities one can estimate the JDOS. Now, one needs
to calculate $T_{\infty}$(I,J;K,L) which can be done by keeping a record of
moves in the form of a matrix, called C-matrix, $C(I,J;K,L)$ for all proposals 
$(I,J) \rightarrow (K,L)$, during the random walk.
Initially, we
set $C(I,J;K,L)=0$ for all I,J and K,L. At infinite
temperature all the proposed moves are accepted so whenever a move is proposed
we update the C-matrix as
\begin {equation}
C(I,J;K,L)=C(I,J;K,L)+1
\end {equation}
\noindent Once the construction of C-matrix is started we never reset it to 
zero, because the C-matrix keeps the detail history of the transitions in
 the system.
The current estimate of the infinite temperature transition probability
($\widetilde{T}_{\infty}$) is
\begin {equation}\label{tes}
\widetilde{T}_{\infty}(I,J;K,L)=\frac{C(I,J;K,L)}{\displaystyle
{\sum_{K}\sum_{L}C(I,J;K,L)}}
\end {equation}
\noindent Where the sum extends over all $K$ and $L$ and the tilde indicates 
the
estimate. From equation (\ref{tw}) and (\ref{tes}) one can determine the
joint density of states. But the equation (\ref{tw}) is an over specified
problem since for a system with $N$ macrostate having $N$ unknown
quantities  $\Omega(I,J)$ there
are $N(N-1)/2$ such equations. To calculate $\Omega(I,J)$ one needs to minimize
the total variance
\begin {equation}\label{var}
\sigma^2_{tot}=\sum_{I,J;K,L}{\displaystyle{\frac{[S(I,J)-S(K,L)+ln\left(\frac{\widetilde{T}_{\infty}(I,J;K,L)}{\widetilde{T}_{\infty}(K,L;I,J)}\right)]^2}{\sigma^{2}_{IJKL}}}}
\end {equation}
\noindent with
\begin {equation}
\sigma^2_{IJKL}=C(I,J;K,L)^{-1}+H(I,J)^{-1}+C(K,L;I,J)^{-1}+H(K,L)^{-1}
\end {equation}
\noindent Here we have used $S(I,J)=ln \Omega(I,J)$ and $H(I,J)=
\sum_{K,L}C(I,J;K,L)$. In order to ensure that the random walker visits
all macrostates in the region of interest one considers a uniform ensemble,  
where all macrostates are equally probable. The probability of occurrence of
a given microstate $(i,j)$ is therefore proportional to the multiplicity
of the macrostate $(I,J)$ where $i\in I$ and $j\in J$.
So the probability of acceptance of a move $(i,j) \rightarrow (k,l)$ is 
given by
\begin{equation}\label{apw}
p(i,j \rightarrow k,l)=min\left(1,\frac{\Omega(I,J)}{\Omega(K,L)}\right)
\end{equation}
\noindent Since the JDOS are not known a priori, the acceptance criteria
can be written as
\begin{equation}\label{att}
p(i,j \rightarrow k,l)=min\left(1,\frac{\widetilde{T}_{\infty}(K,L;I,J)}
{\widetilde{T}_{\infty}(I,J;K,L)}
\right)
\end{equation}

\noindent where we have used equation (\ref{tw}) So using the above acceptance
 probability together with equation
(\ref{tw}) and minimizing the total variance as discussed above, one can
generate the profile of the joint density of states. The TMMC algorithm gives
accurate value of JDOS since it stores and uses the entire history of the
transitions  during the random walk.
But this method has a drawback that the convergence is not guaranteed and
a significant amount of CPU time is necessary.
On the other hand the WL method ensures that all the macrostates are
visited rather efficiently and more quickly. As proposed by Shell and 
co-workers \cite{shell} in the WLTM method we combine the TMMC and WL algorithms
in order to get the benefit of both the algorithms. This is discusses in the
next section.\\

\subsection{\bf Combination of Wang-Landau and Transition Matrix Monte Carlo
algorithm: The WLTM algorithm}
\noindent In the WLTM algorithm, one efficiently combines the WL and TM Monte
Carlo algorithms. In this algorithm, the phase space is sampled via the
acceptance criteria given by equation (\ref{apw}) as in the WL method.
Also $g(I,J)$ and $H(I,J)$ are updated in the same fashion as in the WL
algorithm and in addition of these, a record of transitions between
macrostates is kept in a C-matrix, which is never zeroed during the
simulation. At the end of the simulation, from the knowledge of the
C-matrix and using equation (\ref{tw}) the joint density of states is 
determined by minimizing the total variance.\\

\section{\bf The models used in the present work}
\noindent We have determined the JDOS and other thermodynamic
quantities such as energy, specific heat, order parameter, correlation function
etc. performing two dimensional random walk in $E-\phi$ space using WL and
WLTM algorithms for two continuous lattice spin models.
We give below a brief description of the models and the related quantities
we have measured.\\

\subsection{\bf The 1-d Lebwohl-Lasher model}
\noindent This model is a linear array of three dimensional spins $(d=1, n=3)$ 
interacting
with nearest neighbors via the a potential
\begin {equation}
V_{ij}=-P_2(cos\theta_{ij})
\end {equation}
\noindent Where $P_2$ is the second Legendre polynomial and $\theta_{ij}$ is the
angle between two nearest neighbor spins $i$ and $j$. Spins are headless, i.e.
it has O(3) as well as $Z_2$ symmetry.
This model represents
one dimensional nematic liquid crystal and does not exhibit any finite
temperature order disorder phase transition.\\
\noindent The order parameter is a quantity which describes the amount of
order prevailing in a system. Since in a nematic phase the ordering is the
orientational, the order parameter quantifies the amount of orientational
order present in the system. In a nematic liquid crystal the molecules
are on the average aligned along a particular direction $\bf n$ called the
director. These molecules in general have equal probability of pointing
parallel and anti-parallel to any given direction. If $\bf {\widehat{\eta}(a)}$
is the direction of a molecular axis of a molecule situated at
position $\bf x=a $ then $\bf {-\widehat{\eta}(a)}$ have equal 
contribution as that of
$\bf {\hat{\eta}(a)}$ towards the order parameter. So a vector order parameter 
is
inadequate for the system and a symmetric traceless tensor is used as
the order parameter. This tensor is chosen in such a
way that it is zero in the high temperature isotropic phase and is unity in the
fully ordered phase. The order parameter tensor is given as
\begin {equation}
Q_{ij}=\frac{1}{N}\sum_{i=1}^{N}\left(n_i^an_j^a-\frac{1}{3}\delta_{ij}\right)
\end {equation}
\noindent where $n_i^a$ is the $i^{th}$ component of the unit vector
 $\hat{n}$, which points
along the spin at position {\bf x=a}. $N$ is the number of molecules 
in the system.
The order $S=<q^2>$ prevailing in the system can be written as
\begin {equation}
S=<q^2>=\frac{N}{N-1}<\frac{3}{2}Tr\, Q^2-\frac{1}{N}>
\end {equation}
\noindent The second rank spin-spin correlation function $\rho(r)$
is defined as
\begin{equation}
\rho(r)=<P_2(cos\theta(r))>
\end{equation}
\noindent Where $\theta(r)$ is the angle between two spins separated by 
$r$ lattice
spacing $r$. The order parameter $S$ and $\lim_{r \rightarrow \infty}\rho(r)$
 should
vanish in the thermodynamic limit. However due to the finite size effect in
systems of finite size both quantities have some small non zero value.
The exact expression for the correlation function is given by \cite{romerio} 
\begin{equation}\label{exactll}
\rho_N(r)=\left[\frac{3}{4}\tilde{K}^{-1/2}D^{-1}(\tilde{K}^{1/2})-
\frac{3}{4}\tilde{K}^{-1}-\frac{1}{2}\right]^r
\end{equation}
\noindent where $\tilde{K}=3/2T$. $D$ is Dawson function \cite{dawson},
\begin{equation*}
D(x)=exp[-x^2]\int_0^{x}du\,\, exp[u^2].
\end{equation*}
\subsection{\bf The 2-d XY model}
\noindent In this model planar spins placed at the sites of a planar
square lattice interact with nearest neighbours via a potential,
\begin{equation}\label{vXY}
V(\theta_{ij})=2\left\{1-\left[\cos^2(\theta_{ij}/2)\right]\right\}
\end{equation}
\noindent where $\theta_{ij}$ is the angle between the nearest neighbours
$i, j$. [This particular form of the interaction, rather than the more
conventional $-\cos(\theta_{ij})$ form, was chosen by Domany {\it{et. al}}
\cite{domany}
to enable them to modify the shape of the potential easily, which
led to what is now known as the modified XY-model ]. The XY-model is
known to exhibit a quasi-long-range-order disorder
transition which is mediated
by unbinding of topological defects as has been described in the seminal
work of Kosterlitz and Thoules \cite{KT, K}. The XY-model has also been
the subject
of extensive MC simulation over last few decades and some of the
recent results may be found in \cite{palma}.\\
\noindent In this model, the orientational order parameter is defined
as follows: let $\bf n$ be the unit vector (called the director) 
in the direction
of maximum
order prevailing in the system and $\bf s$ be the spin vector of unit
magnitude then the order parameter $S$ is given by
\begin{equation} \label{opxy}
S=<{\bf {n}\cdot \bf {s}>}\\
 =<cos \, \phi>
\end{equation}
\noindent where $\phi$ is the angle between the director and the spin. The spin-
spin correlation function for the 2d-XY model is defined as
\begin{equation}
\rho(r)=<cos\,\theta(r)>
\end{equation}
\noindent Where $\theta(r)$ is the angle between two spins separated by $r$
lattice spacing.\\
\section {\bf Computational details:}
\noindent Two dimensional random walk has been performed in $E-S$ and $E-\rho$
space for one dimensional Lebwohl-Lasher model and two dimensional XY model.
Since both of these models are continuous, we discretize the system by
dividing the whole energy range as well as the order parameter (or correlation
function) range into a number of bins having widths $d_e$ and 
$d_\phi$ respectively. We simulated 2d XY system for lattice sizes
$5$, $10$, $15$ and $20$ and 1d LL system for lattice sizes $80$, $160$ and
$220$. The 2d XY system can have energy between $0$ and $2N^2$ but we
simulated the system for the energy range $3.0$ to $2N^2$ to avoid trapping 
of the random walker in
low energy states as these are scarcely visited during the simulation.
Similarly the 1d LL system can have energy between $-N$ and $N/2$ but we
simulated the system for the energy range $-N$, $0$ with a small energy cut 
near
the ground state.
We have chosen $d_e$ to be $0.1$ for the 1d LL model and
$0.2$ for the XY model.
$d_\phi$ was chosen to be $0.01$ for both the models while performing 
random walk
in $E-S$ space as well as in the $E-\rho$ space. In the 1d LL model, 
each spin have three
components ($l_i$ with $i=1, 2, 3$) specifying the three direction 
cosines of the
spin. We have taken a random initial configuration and a new configuration
is generated by rotating any one of the spins randomly using the prescription
$l_i=l_i+p*r_i$ (for i=1,2,3), where $r_i$ is a random number between -1 
and 1. In the 2d XY model, each spin is specified by two
direction cosines and new configurations are generated by rotating the spin
in the same manner as above. Initially, we do not have any prior knowledge
about the DOS, so we set $g(I,J)=0$ for all values of 
the macrostates $I$ and
$J$. A new microstate $(k,l)$ is generated from the old one $(i,j)$ by rotating
one spin at a time and the acceptance probability $p(i,j;k,l)$ is given 
by equation
(\ref{aprob}). When the new state lies in the same macrostate as the old one
the state is accepted. A histogram count is recorded in an array $H(I,J)$
and whenever a state is proposed we
update the C-matrix as $C(I,J;K,L)=C(I,J;K,L)+1$. The random walk is continued
till the histogram becomes flat (say $80\%$). We point out that
in the case of 2-d random walk all the bins are not visited uniformly
as this would need an enormous number of Monte Carlo Sweeps (MCS). 
We first sampled
the system for about $10^6$ MCS which is called the 'pre-production run'. During
the pre-production run we marked by $'1'$ the bins which are visited at
least $80\%$ of its average value and by $'0'$ those which are visited less than
$80\%$ of the average value or are not visited at all. During the 'production 
run' we checked
the histogram flatness only for those bins which were marked $'1'$
. The average histogram is calculated by considering only those bins which are
visited at least once discarding the bins which are not visited at all.
Some bins marked $'0'$ may qualify for the  mark $'1'$ during the production
runs.
When the histogram gets flat
the modification factor is changed as 
$lnf \rightarrow lnf/2$ and
the histogram count is reset to zero but the C-matrix is never reset to zero.
The iteration is continued with the new modification factor and the same
procedure is repeated until the modification factor becomes as small as
$10^{-7}$ for LL model and $10^{-5}$ for XY model. At the end of the simulation
the JDOS is calculated from the knowledge of the C-matrix by
minimizing the variance in equation (\ref {var}). This variance is minimized
using the Dowhill-Simplex method \cite{downhill}.

\noindent It may be noted that for the continuous system there are a large
number of macrostates $I$ and $J$. As a result, the number of elements
in the four dimensional C-matrix is enormously large.
To simulate the system a huge amount of computer storage (RAM)
is required which is beyond of our computer resources.
But most of the elements of C-matrix
are zero since a transition from a given macrostate $(I,J)$ to all macrostates
$(K,L)$ is not possible. In general it is found that for the transition
$(I,J) \rightarrow (K,L)$, the possible nonzero values of $K$ and $L$ lie
within the range $I-n_1$ to $I+n_1$ and $J-n_2$ to $J+n_2$ respectively,
where $n_1$ and $n_2$ are integers.
Some extra conditions are imposed for the low values of $I$ and $J$.
In our case the values of $n_1$ and $n_2$ are 5 and 1 respectively.
So to avoid the problem with storage we transformed $K$ to $K'=K-I+n_1$ and
$L$ to $L'=L-J+n_2$ and whenever the original values of $K$ and $L$ are
necessary we make the reverse transformations.\\
The JDOS is obtained by minimizing the variance using the Downhill Simplex
method \cite{downhill}.
The variance is a function of more than one independent variable; in fact it
is a function of a large number of independent variables.
 It is also very difficult
to minimize the multivariable function of such a large number of 
unknown variables
using this method. For example, to minimize a function of $N$ variables (which
constitutes a point in an N-dimensional space) it requires 
$N+1$ points in that space.
One of these points is chosen as the starting point $P_0$ and other points are
obtained by $P_i=P_0 +\lambda e_i$, where $\lambda$ is a constant and
is taken to be $0.1\%$ of the value of $P_0$, 
and the
$e_i$ s are $N$ unit vectors. In this method we need a two dimensional
matrix ($p$) of extents $N+1$ and $N$. For a continuous model,
 the values of $N$ is
very large consequently the $p$ matrix becomes very large. We had to divide
the entire two dimensional
surface $g(I,J)$ into a large number of smaller segments. 
The energy-order parameter
surface $g(I,J)$ of LL model was divided into $79$, $159$, $218$ segments
 for the lattice
sizes $80$, $160$ and  $220$ respectively. Each segment consists of 
$100$ order parameter
bins and $10$ energy bins. Similarly, the energy-correlation function
surface of LL model is divided into the same number of segments as before each
having $140$ correlation function bins and $10$ energy bins.
We divided the energy-order parameter surface of XY model into $224$, $984$,
$2132$ and $3980$ segments for lattice sizes $5$, $10$, $15$ and $20$
respectively. Each segment contains $100$ order parameter bins and $14$,
$12$, $12$, $10$ energy bins for lattice sizes $5$, $10$, $15$ and $20$
respectively. Similarly the energy correlation function surface of 
the XY model was also
divided into the same number of segments. Each of the segments contains
$100$ correlation function bins and $10$ energy bins.
Minimization of the variance given by equation (\ref{var}) was carried out
separately over each segment using the
Downhill Simplex method
and the JDOS for the entire surface is obtained by connecting the JDOS of each
segment obtained by the above mentioned minimizing method. 
With the above procedure
we were able to calculate the JDOS of energy-order parameter or 
energy-correlation
function by minimizing the variance as stated.\\
\begin{figure}[ht]
\begin{center}
{\includegraphics[width=0.9\linewidth]{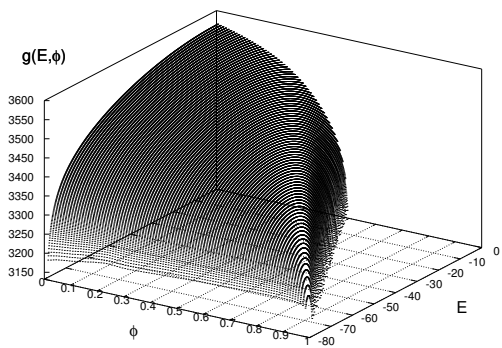}}
\end{center}
\caption{The logarithm of JDOS for the $L=80$ 1-d LL model obtained
from the WLTM method is plotted against the energy and order parameter.
}
\label{jdos.eps}
\end{figure}
\begin{figure}[ht]
\begin{center}
{\includegraphics[width=0.9\linewidth]{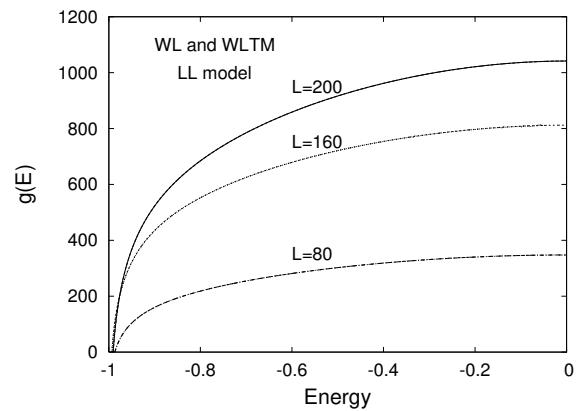}}
\end{center}
\caption{Plot of the logarithm of density of states, $g(E)$ against 
system energy per
particle for 1-d LL model of lattice size $L=80$, $160$ and $220$ obtained
using WL and WLTM algorithms.
}
\label{dosll.eps}
\end{figure}
\begin{figure}[ht]
\begin{center}
{\includegraphics[width=0.9\linewidth]{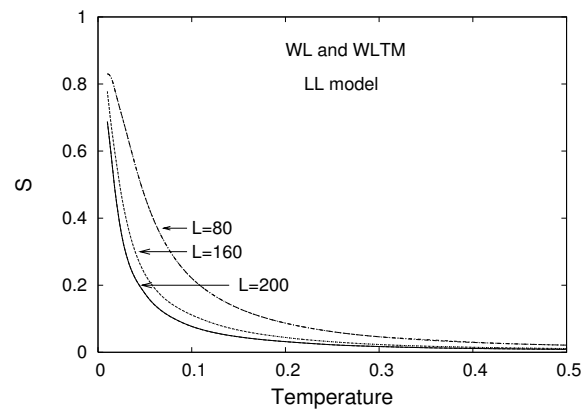}}
\end{center}
\caption{The variation of orientational order parameter $(S)$ fo the 
1-d LL model
for different lattice sizes is plotted with temperature, obtained from both
WL and WLTM algorithms.
}
\label{orderll.eps}
\end{figure}
\begin{figure}[ht]
\begin{center}
{\includegraphics[width=0.9\linewidth]{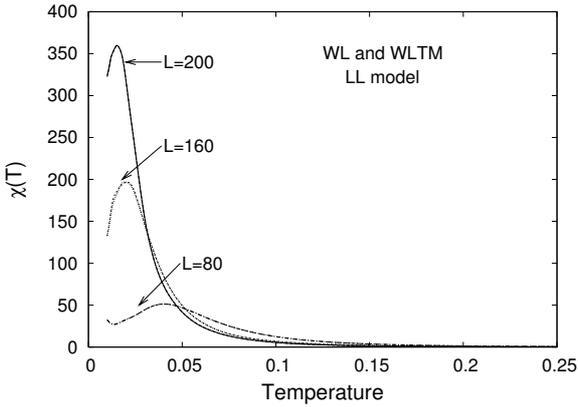}}
\end{center}
\caption{The Susceptibility is shown as a function of temperature for the 1-d LL
model of different lattice sizes. Data obtained from WL algorithm as well as
from WLTM algorithm are plotted in this graph.
}
\label{susll.eps}
\end{figure}
\begin{figure}[ht]
\begin{center}
{\includegraphics[width=0.9\linewidth]{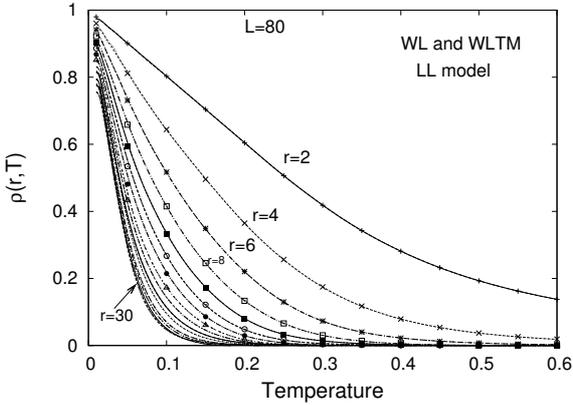}}
\end{center}
\caption{ Variation of correlation function $\rho(r,T)$ with temperature
for lattice size $L=80$ of 1-d LL model. $15$ different values of $r$
taken $2$, $4$, $6$, $8$, $10$, $12$, $14$, $16$, $18$, $20$, $22$, $24$,
$26$, $28$ and $30$. The top most curve is for $r=2$ and the lower curves
are for other values of $r$ given in the sequence above and in ascending
order of $r$. The curves are the results obtained from the JDOS
obtained using WL and WLTM algorithms. Comparison has been made with the
exact results (the dots) obtained from \cite{romerio} using equation
\ref{exactll} for $r=2$, $4$, $6$, $8$, $10$, $12$, $14$ and $16$.
}
\label{corll.eps}
\end{figure}
\begin{figure}[ht]
\begin{center}
{\includegraphics[width=0.9\linewidth]{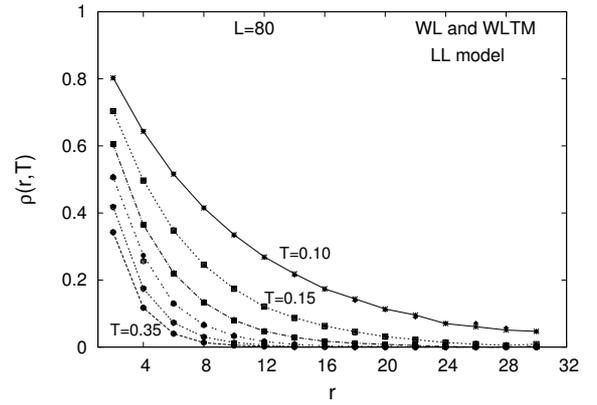}}
\end{center}
\caption{Correlation function as a function of lattice spacing $r$ for seven
different temperatures. Both WL and WLTM algorithms have been used.
Comparison has been made with the exact results (the dots) of \cite{romerio}
obtained using equation \ref{exactll}.
}
\label{rhorll.eps}
\end{figure}
\begin{figure}[ht]
\begin{center}
{\includegraphics[width=0.9\linewidth]{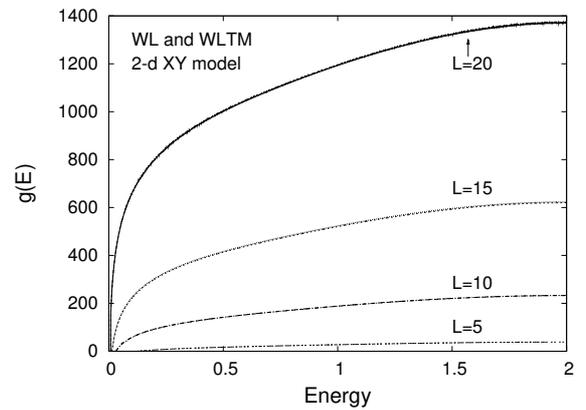}}
\end{center}
\caption{ Logarithm of density of states vs energy per particle for the 2-d XY
model of sizes $5 \times 5$, $10 \times 10$, $15 \times 15$ and $20\times 20$.
The results obtained from both WL and WLTM algorithm are plotted.
}
\label{dosxy.eps}
\end{figure}
\begin{figure}[ht]
\begin{center}
{\includegraphics[width=0.9\linewidth]{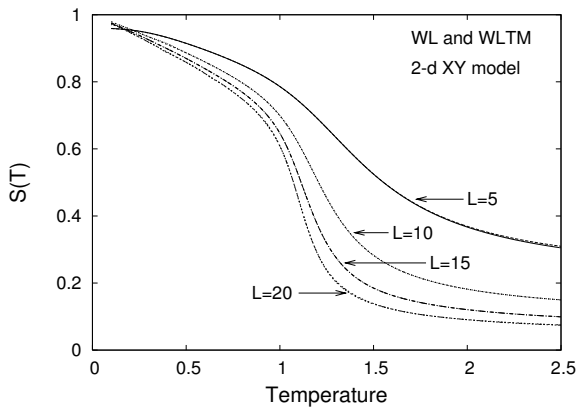}}
\end{center}
\caption{ The orientational order parameter is shown against temperature $T$
for 2-d XY model of different lattice sizes using both WL and WLTM algorithms.
}
\label{orderxy.eps}
\end{figure}
\begin{figure}[ht]
\begin{center}
{\includegraphics[width=0.9\linewidth]{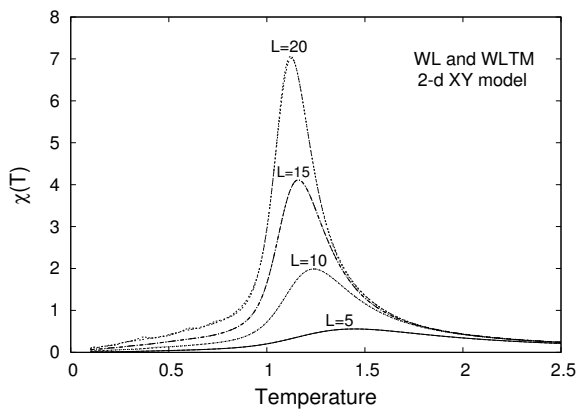}}
\end{center}
\caption{ The susceptibility of 2-d XY model for four liner lattice sizes
$L=5$, $10$, $15$ and $20$ against temperature is plotted. All the data are
from the results of two dimensional random walk in $E-S$ space using WL
and WLTM algorithm. The height of the peak increases with the increase in
system size. It is also noted that the position of peak shifts towards
lower temperature with the increase of system size.
}
\label{orderxy.eps}
\end{figure}
\begin{figure}[ht]
\begin{center}
{\includegraphics[width=0.9\linewidth]{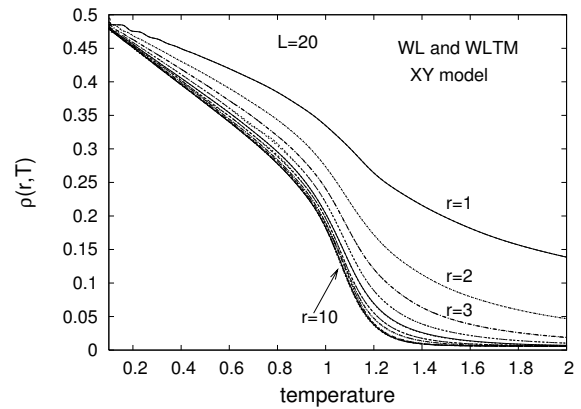}}
\end{center}
\caption{ Variation of correlation function $\rho(r,T)$ with temperature
for lattice size $L=20$ of the 2-d XY model. $10$ different values of $r$
taken $1$, $2$, $3$, $4$, $5$, $6$, $7$, $8$, $9$ and $10$.
The top most curve is for $r=1$ and the lower curves
are for other values of $r$ given in the sequence above and in ascending
values of $r$. The curves are the results we obtained from joint density
of states obtained using the both WL and WLTM algorithms.
}
\label{corxy.eps}
\end{figure}
\begin{figure}[ht]
\begin{center}
{\includegraphics[width=0.9\linewidth]{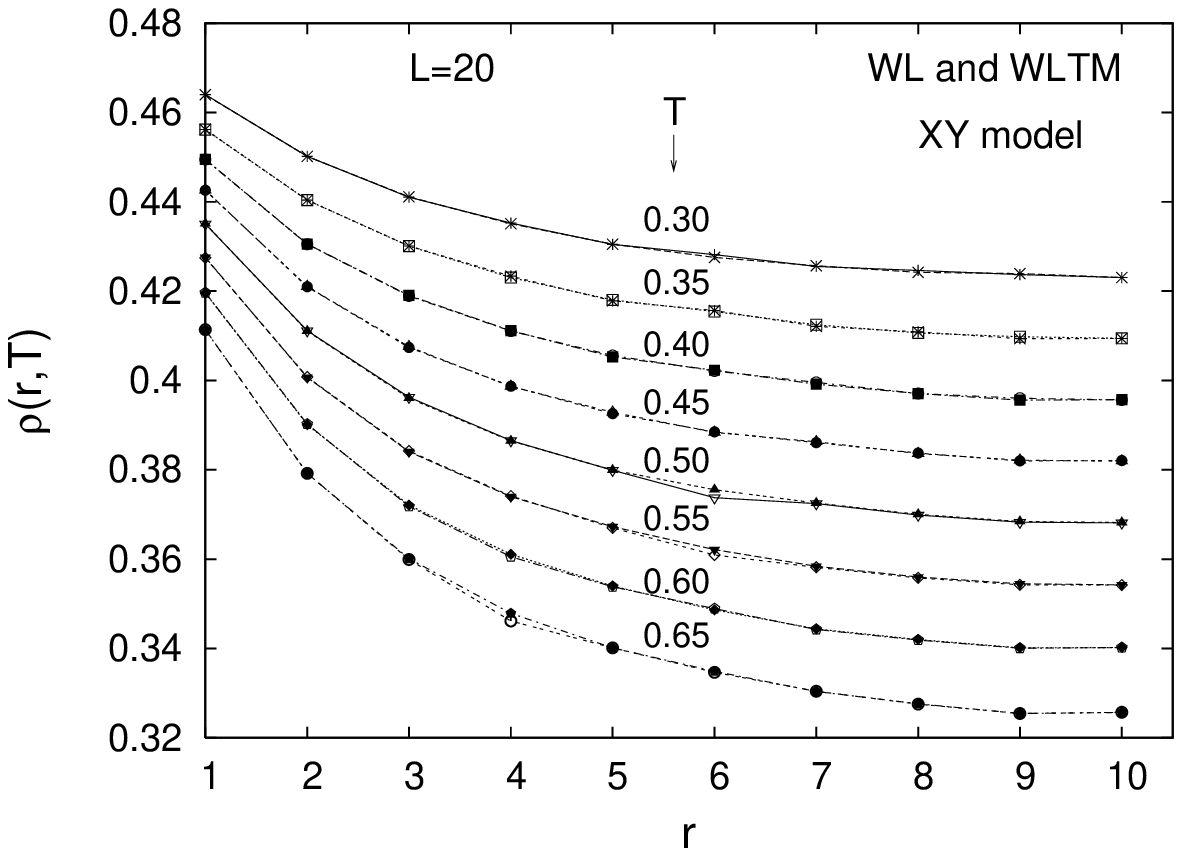}}
\end{center}
\caption{Correlation function as a function of lattice spacing $r$ with eight
different temperatures as parameter is plotted. Both the WL and WLTM algorithms
were used.
}
\label{rhorxy.eps}
\end{figure}
\section{\bf Results and discussion}
\subsection{The 1-d LL model:}
\noindent In this model simulation was carried out for systems having
linear dimension $L=80$, $160$ and $220$. The JDOS obtained by using
the WLTM algorithm is shown in figure 1 as a function of energy and
order parameter for $L=80$.
In figure 2 we have plotted the DOS (actually its logarithm, $g(E)$) for
the three lattices as a function of energy per particle obtained from the
JDOS of figure 1 by using both WL and WLTM methods.
From the knowledge of $g(E)$ one can calculate the average energy, 
specific
heat etc. To determine the order parameter $S(T)$, one needs to perform 
2d random
walk in energy-order parameter space. Figure 3 shows the order parameter
as a function of temperature for the 1d LL model for different lattice sizes
obtained using WL and WLTM algorithms.
From the figure it is clear that S decreases rapidly with temperature and
with increase in system size. This is expected since
the 1d LL model does not possess a true long range order at any finite 
temperature and the non-zero values of $S$ obtained is 
due to the finite size effect. Consequently the 
1d LL model does not exhibit any finite temperature order-disorder phase
transition. In figure 4, susceptibility is plotted against temperature for
different system sizes. The peaks become sharp with the increase
of system size. This also supports the absence of order-disorder phase
transition. \\
The correlation function $\rho(r,T)$ has been plotted against
temperature with $r$, the spacing between spins, in figure 5
for lattice size L=80 and 15 different values of $r$ ranging from
$2$ to $30$. All data are obtained from 2-d random walk in $E-\rho$ space
using both WL and WLTM algorithms. It may be noted that,
we had to run one simulation for each value of $r$. With increase in
temperature, the correlation between spins for a given value of $r$ decreases.
Again $\rho(r,T)$ is plotted as a function of lattice spacing $r$ for seven
different temperatures in figure 6. $\rho(r,T)$ decreases very
rapidly with lattice
spacing {\it i.e.} spins which are a large distance apart are uncorrelated.
This is expected since in the thermodynamic limit,
$\lim_{r \rightarrow \infty}\rho(r)=0$. The finite values of correlation
function that appears in the figures is due to the finite size effect.\\

\subsection {\bf The 2-d XY model}
\noindent In this model the simulation has been carried out for lattice
sizes $5 \times 5$, $10 \times 10$, $15 \times 15$ and $20 \times 20$ using
both WL and WLTM algorithms. Two dimensional random walk in two different
spaces ($E-S$ and $E-\rho$) was performed to determine the order parameter,
susceptibility and correlation function apart from average energy, specific
heat etc which can be determined using 1-d random walk. The minimum energy
for all lattice sizes was $3$. The upper limit of energies of these system
sizes over which the simulation has been carried out were $50$, $200$,
$450$ and $800$. We deleted a small region at the lower end of energy to
overcome the trapping of the random walker. The order parameter
(correlation function) can have values between $0$ and $1$ ($-0.5$ and
$+0.5$).
The whole energy and order
parameter (correlation function) range is
divided into a large number of bins of width $d_e=0.2$ and $d_\phi=0.01$
respectively.\\
\noindent In figure 7, the density of states obtained from the 2-d random
walk for 2-d XY model for different lattice sizes have been plotted against
the energy per particle. We have used both WL and WLTM algorithms to obtain
the data presented in figure 7. The order parameter ($S(T)$) for 2-d XY model
of linear lattice sizes $5$, $10$, $15$ and $20$ are plotted with temperature
in figure 8. The system is known to
possess no true long range order and a quasi-long-range-order disorder
transition takes place due to unbinding of topological defects.
Susceptibility of 2d XY
model is also plotted as a function of temperature for four lattice sizes in
figure 9. It is observed that the height of the susceptibility peak 
increases with
the increase of system sizes and also the position of susceptibility peak
is shifted towards the lower temperature with the increase of system size.
In figure 10 we have plotted the correlation function $\rho(r,T)$ against
temperature for ten values of $r$ for the XY-model. These were obtained
from the JDOS computed by using WL and WLTM methods. The same data
is depicted in a different way in figure 10, where we have plotted $\rho(r, T)$
against $r$ for eight different values of temperature.\\
\section{Conclusion}
\noindent We have presented the results of Monte Carlo simulation performed
in the 1-d LL model and 2-d XY-model for the evaluation of joint density
of states using the WL and WLTM algorithms. Agreement of the statistical
averages of different quantities obtained by using the two algorithms is
excellent. Calculation of JDOS for continuous spin models has been done
earlier \cite{kisor}.
We have demonstrated in this work that, although computationally tedious,
 it is possible to use the WLTM method for the evaluation of the JDOS for
a continuous spin model. This method may prove to be useful for future
researchers who will need to generate the JDOS in a discrete or coninuous
spin model.\\
\section{Acknowledgment}
\noindent We acknowledge the receipt of a research grant No.
03(1071)/06/EMR-II from Council of Scientific and Industrial Research (CSIR),
India which helped us to procure the IBM x226 servers. One of the authors (SB)
gratefully acknowledges CSIR, India, for the financial support.

\end {document}